\documentclass[aip,pop]{revtex4-1}

\usepackage{graphics}
\usepackage{graphicx}

\begin{document}

\title{Broad-band polarization-independent absorption of electromagnetic waves by an overdense plasma}
\author{Y.P. Bliokh}
\affiliation{Department of Physics, Technion, 32000 Haifa, Israel}
\author{Yu.L. Brodsky}
\affiliation{Department of Physics, Technion, 32000 Haifa, Israel}
\author{Kh.B. Chashka}
\affiliation{Department of Physics, Technion, 32000 Haifa, Israel}
\author{J. Felsteiner}
\affiliation{Department of Physics, Technion, 32000 Haifa, Israel}
\author{Ya.Z. Slutsker}
\affiliation{Department of Physics, Technion, 32000 Haifa, Israel}

\begin{abstract}
Surface plasmon-polaritons can be efficiently excited on a
plasma-vacuum interface by an electromagnetic wave when a
subwavelength diffraction grating is placed in front of the plasma
boundary. The excitation efficiency depends strongly on the wave
frequency (or plasma density, when the frequency is fixed) and
polarization. We show both experimentally and theoretically that
this sensitivity can be essentially suppressed. A non-zero angle
of incidence and an axially-symmetric diffraction grating ensure
near-total absorption of the incident wave in a broad range of
wave frequencies (or plasma densities, when the frequency is
fixed). Direct detection of surface plasmon-polaritons has been
achieved for the first time using a miniature antenna imbedded in
the plasma. A new absorption mechanism which is not associated
with surface plasma waves excitation is revealed.

\end{abstract}


\maketitle

\section{Introduction}
Surface plasma waves -- surface plasmon-polaritons (SPP) --
discovered more than fifty years ago \cite{Ritchie}, are now
attracting an increased attention due to their numerous realized
or promising applications. Intensive investigations of SPP in the
optical frequency range have led to appearance of a new field of
physics -- plasmonics \cite{Zia, Ozbay}. Resonant excitation of
SPP is responsible for such effects as the extraordinary
transparency of optically thick perforated metal films
\cite{Ebbesen},  resonant transmission of electromagnetic waves
through overdense-plasma structures \cite{Sternberg}, frustrated
total internal reflection \cite{Otto, Kretschmann}, and
superresolution of Veselago-Pendry's ``perfect'' lens
\cite{Veselago, Pendry, Ruppin}.

A freely propagating electromagnetic wave cannot excite SPP
because of the mismatch between their phase velocities along the
medium-vacuum interface. The coupling between the electromagnetic
wave and SPP can be achieved either in the total internal
reflection configuration \cite{Otto, Kretschmann} or due to the
interface corrugation (periodic, random corrugation, film
perforation, etc.). It is worth to notice that it is a question of
propagating surface waves. Localized (non-propagating)
plasmon-polaritons can be excited by an electromagnetic wave on
the surface of subwavelength-size objects \cite{Pillai, Teperik}.

Both these methods are acceptable when SPP are excited at a metal
interface, and are difficult for realization for SPP excitation at
a plasma interface. In the latter case the most convenient is the
use of a subwavelength diffraction grating as a matching element
\cite{Bliokh_1, Bliokh_opt, Tripathi}. It has been shown both
theoretically and experimentally in \cite{Bliokh_1} that the SPP
excitation with the use of the diffraction grating properly placed
in front of the plasma surface leads to  total absorption of the
incident electromagnetic wave by an overdense plasma, i.e. by a
plasma whose Langmuir frequency, $\omega_p$, exceeds the wave
frequency $\omega$. This effect has been observed in a plasma with
other parameters \cite{Plasma, Chen} and in a solid state plasma
\cite{Solid}.

Due to its resonant nature, the SPP excitation is possible in a
relatively narrow range of parameters (plasma density, wave
frequency and polarization, angle of incidence, etc.). This
property may be utilized in a set of applications such as
frequency and angle-resolved filters, for example. On the
contrary, other problems like, for example, plasma heating,
development of a non-reflecting coating, and enhancement of
photovoltaic cells efficiency, need broadening of the parameters
region where SPP are effectively excited. Here, we propose and
study both theoretically and experimentally some methods of the
parameters region broadening. The operation in the microwave
frequency range allowed us to achieve (for the first time, as far
as we know) direct detection of surface plasma waves by a
miniature antenna imbedded in the plasma. It is also shown that
the total absorption of an electromagnetic wave by an overdense
plasma can also be achieved without SPP excitation.

\section{Physical background}

The phase velocity of SPP propagating along the plasma-vacuum
interface is smaller than the speed of light $c$, whereas the
projection of the incident wave phase velocity on the interface
plane, $c/\sin\theta$ (where $\theta$ is the angle of incidence),
exceeds the speed of light. This mismatch makes impossible direct
excitation of SPP by a freely propagating electromagnetic wave.

When a diffraction grating is placed in front of the plasma
surface, the scattered electromagnetic field becomes richer by new
spatial harmonics whose wave vectors are
$\mathbf{k}_\ell=\mathbf{k}_0+\ell\mathbf{k}_g$, where
$\mathbf{k}_g$ is the grating reciprocal vector, $\mathbf{k}_0$ is
the incident wave vector, $k_0=\omega/c$, and
$\ell=\pm1,\pm2,\ldots$. Some of these harmonics can be in
resonance with SPP and excite them when the following resonance
condition is satisfied:
\begin{equation}\label{eq1}
k_p(\omega,n_p)\equiv k_0\left({\varepsilon\over
1+\varepsilon}\right)^{1/2}=k_{\ell,t},
\end{equation}
where $n_p$ is the plasma density,
$\varepsilon=1-\omega_p^2/\omega^2$ is the plasma permittivity,
$\omega_p=\sqrt{4\pi e^2n_p/m}$ is the plasma Langmuir frequency,
$k_p(\omega,n_p)$ is the wave vector of SPP whose frequency is
$\omega$, and $k_{\ell,t}$ is the tangential component of wave
vector $\mathbf{k}_\ell$. If the incident wave frequency and the
angle of incidence are fixed, equation~(\ref{eq1}) determines the
resonant plasma density $n_\ell$.  The resonant plasma densities
$n_{\pm1} $(below we will consider the strongest, $\ell=\pm 1$,
harmonics only) are given by the following relation:
\begin{equation}\label{eq1a}
\eta_{\pm1}\equiv {n_{\pm1}\over n_c}={2(k_g\pm
k_0\sin\theta)^2-k_0^2\over (k_g\pm k_0\sin\theta)^2-k_0^2},
\end{equation}
where $n_c$ is the critical density defined by the condition
$\omega_p(n_c)=\omega$.

Besides the resonance condition (\ref{eq1}), the incident wave
must be properly polarized. Decomposing an arbitrarily polarized
incident wave into two linearly polarized waves  whose magnetic
fields have tangential projections directed along and
perpendicular to the grating strips (wave-1 and wave-2
thereafter), one can see that only wave-1 being scattered on the
grating excites SPP when the resonant condition (\ref{eq1}) is
satisfied.

The SPP electromagnetic fields are concentrated in the vicinity of
the plasma-vacuum interface and can be much stronger than the
incident wave fields. Due to this enhancement of the
electromagnetic energy density, even small dissipation in the
plasma can lead to a significant absorption of the incident wave
energy. In this case electromagnetic wave reflection from the
grating-plasma system is small or even equal to zero. The
reflection coefficient $R$ drops to zero, $R=0$, under certain
conditions which can be easily understood considering a surface
wave as a resonator \cite{Bliokh_1, Bliokh_2}. The Q-factor of
this resonator, as the Q-factor of any resonator, is determined by
two factors: dissipative and radiative losses. The first one is
connected with all kinds of dissipative processes (collisions in
plasma, ohmic losses in the grating, etc.). The second one is
determined by the efficiency of the transformation of the surface
(evanescent) wave into the freely propagating wave under
scattering on the grating. These losses are characterized by
corresponding Q-factors, ${\rm Q}_{diss}$ and ${\rm Q}_{rad}$.
When both Q-factors are equal one another, ${\rm Q}_{diss}={\rm
Q}_{rad}$, the reflection coefficient vanishes. This phenomenon is
well-known in microwave electronics as  \textit{critical coupling}
(see, e.g., \cite{Slater, Adler}).

Due to the resonant nature of this phenomenon, the reflection is
small either in a narrow frequency band $\Delta\omega$ near the
resonant frequency (when the plasma density and the angle of
incidence are fixed) or in a narrow region $\Delta n_p$ of the
plasma density values (when the wave frequency and the angle of
incidence are fixed). Using equation~(\ref{eq1}) it is easy to
calculate $\Delta\omega$ when $\Delta n_p$ is known and vice
versa. Taking into account the experimental conditions (fixed
$\omega$ and $\theta$, and varying $n_p$) we will consider $\Delta
n_p$ as the resonance width.

Thus, the SPP excitation is sensitive to the deviation of the
system parameters  from their resonant values as well as to the
relative polarization of the incident wave with respect to the
grating. There is a set of problems where this sensitivity plays a
positive role and can be utilized, for instance, as frequency,
angular or polarization filters. On the other hand, the set of
applications such as electromagnetic cloaking \cite{Alu, Alitalo},
solar cells \cite{Pillai, Akimov}, etc., needs suppression of this
sensitivity.

The resonance can be broadened in the following way. Under a
normal incidence on a 1D grating (grating that is characterized by
one vector $\mathbf{k}_g$) the electromagnetic wave excites two
equal SPPs propagating in opposite directions. The resonant plasma
density $n_{res}$ is the same for both SPPs. Strong absorption is
observed in a certain region $\Delta n_p$ around $n_{res}$. If the
angle of incidence $\theta$ deviates from zero, two SPPs, whose
wave vectors are $k_{\pm1}=k_0\sin\theta\pm k_g$, can be excited.
Now there are two different resonant plasma densities, $n_{\pm1}$,
and the reflection is suppressed in the regions $\Delta n_{\pm1}$
around them (when the plasma density is fixed, there are two
frequency bands for which reflectivity is suppressed). When the
difference between the two resonant densities is of the order of
the widths $\Delta n_{\pm1}$, the resonances overlap and the
resulting resonant region is broadened.

The above-mentioned sensitivity of the SPP excitation to the
incident wave polarization can be suppressed when the 1D grating
is replaced by a 2D grating. It can be a periodic grating which is
characterized by two mutually orthogonal reciprocal vectors
${\mathbf k_{g1}}$ and ${\mathbf k_{g2}}$. If the lengths of these
vectors are close to one another, $k_{g1}\simeq k_{g2}$, two
components of the incident wave, wave-1 and wave-2, excite surface
plasmon-polaritons simultaneously and the reflection coefficient
can be small independently of the wave polarization \cite{Popov}.
More complex aperiodic gratings characterized by many reciprocal
vectors can also be used \cite{Matsui}. When the angle of
incidence is not equal to zero but small, $\theta\ll 1$, every
resonance splits into two ones. Overlapping of these resonances
can lead to the broadening of the parameters region where
reflection is suppressed.

\begin{figure}[tbh]
\centering {\scalebox{0.5}{\includegraphics{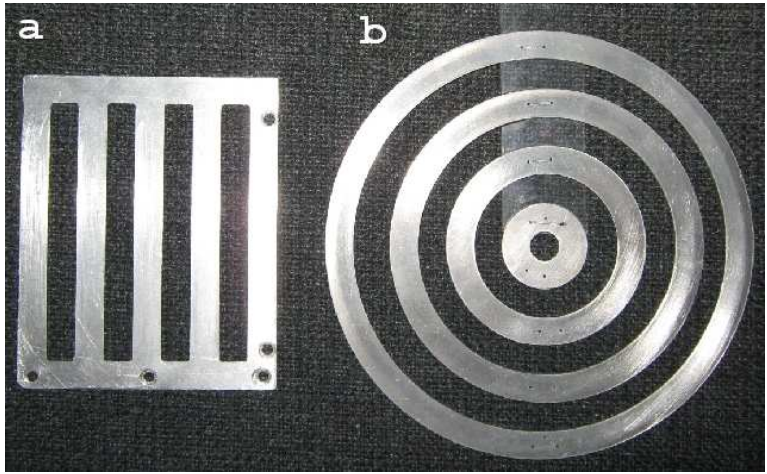}}
\caption{1D and 2D circular gratings.\label{Fig1a}}}
\end{figure}

In order to eliminate \textit{a priori}  the dependence on the
polarization for normally incident wave we suggest a
\textit{circular} diffraction grating as depicted in
figure~\ref{Fig1a}. A qualitative argument for this choice is the
following. When the grating radius $R_g$ is large as compared with
the wavelength $\lambda$, $R_g\gg\lambda$, the grating can be
characterized by a \textit{local} grating vector
$\mathbf{k}_g(\varphi)$, whose length is constant and its
direction depends on the polar angle $\varphi$. The local
efficiency $q$ of the surface wave excitation by a linearly
polarized incident wave depends on the angle $\psi$ between the
local grating vector $\mathbf{k}_g(\varphi)$ and the tangential
component of the wave electric field, $q\sim|\cos\psi|$. There are
regions where the efficiency $q$ is small or even equal to zero.
Reflection from these regions is large and all the energy of the
incident wave cannot be absorbed by the plasma. However, when the
grating radius does not exceed several wavelengths, this
conclusion is not valid. The reason is that the conception of the
local characteristics of the grating is applicable to regions
whose dimension is comparable with the wavelength and is small as
compared with the grating dimension. When the grating dimension is
not so large, the grating surface cannot be separated into regions
with low and high efficiency $q$. In this case the whole grating
is characterized by a unified efficiency $q$ and a wave beam
focused on the grating can be totally absorbed independently of
polarization.  When the angle of incidence is not equal to zero
but small, $\theta\ll 1$, the dependence on the wave polarization
is strongly suppressed. It is possible to suppose that a wide wave
beam can be absorbed using a superlattice composed of circular
gratings.

All the above-mentioned statements related to the resonance
broadening and polarization-independent efficiency of the
electromagnetic wave absorption have been verified experimentally
and results of the experiments are presented in the following
sections.

\section{Experimental results}

\subsection{Experimental setup}

\begin{figure}[tbh]
\centering \scalebox{0.6}{\includegraphics{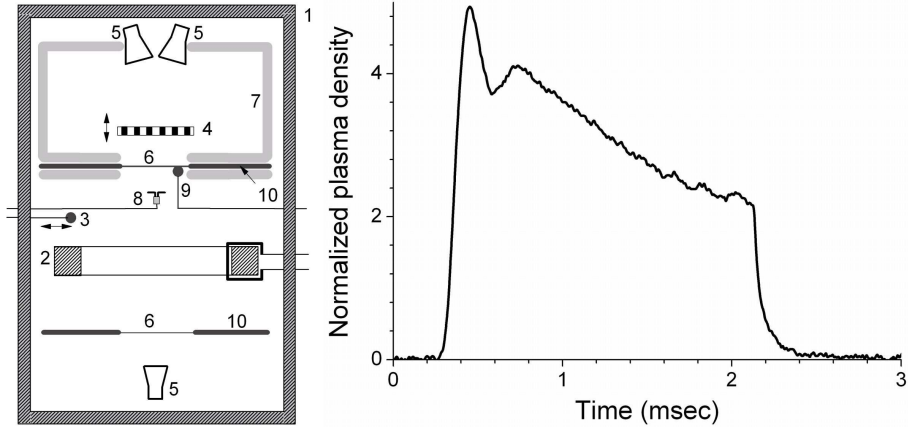}}
\caption{Left -- experimental setup. 1 -- vacuum chamber, 2 --
ferromagnetic core, 3 -- movable Langmuir probe, 4 -- diffraction
grating, 5 -- microwave horns (transmitter and receiver), 6 --
microwave windows, 7 -- microwave absorber, 8 --  microwave
antenna, 9 -- surface Langmuir probe, 10 -- plasma forming plates.
Right -- waveform of the normalized plasma density,
$\eta=n_p/n_c$, variation.} \label{Fig1}
\end{figure}

A sketch of the experimental setup is shown in figure~\ref{Fig1}.
A uniform 20 cm diameter plasma ``dish'' was formed by a
ferromagnetic inductively coupled plasma source, driven by a
pulsed (2 ms) 10 kW rf oscillator (not shown in the sketch). The
plasma appeared between planar plasma forming plates (PFP). The
plasma parameters (density $n_p$ and electron temperature $T_e$)
were measured by movable and surface Langmuir probes. All the
probe measurements were verified by the microwave cut-off method
which provides low tolerance ($\sim3$\%) of the density
measurements. For more details on the plasma source and plasma
parameter measurements see \cite{Plasma_Ex}. The experiments were
carried out in a vacuum chamber of 1.2 m length and 0.7 m diameter
filled with Xe gas in the pressure range (0.7 - 2.4) mTorr. The
electron  temperature $T_e$ was found to be (2 - 4) eV. As a
microwave source we used a klystron oscillator whose operating
frequency $f_0$ was 8.45 GHz (wavelength $\lambda=3.55$ cm). The
plasma density $n_p$ in the vicinity of the PFP decreases
monotonically during the pulse from $n_p=4n_c$ to $n_p=2n_c$. The
characteristic waveform of the plasma density variation is
depicted in figure~\ref{Fig1}. In the center of each PFP there was
made a rectangular ($15\times 15$ cm) window of thin (0.1 mm)
plastic film providing negligible microwave reflection. The
transmitting microwave horn and the one receiving the reflected
signal were placed tightly together in the outer side of one of
the PFP at 17 cm distance from it by such a way that the main lobe
would cover the window. The small angle between these two horns
($\pm 8^\circ$ from the central axis) was adjusted so as to
maximize the microwave signal reflected from the plasma. This PFP
as well as the chamber walls in this chamber section were covered
by microwave absorber to prevent parasitic reflections. A single
horn, placed in the opposite section of the chamber was used for
microwave calibration of the plasma density. In between the above
pair of horns and the window we placed a thin movable metal
grating (see figure~\ref{Fig1a}a,b) which served to excite SPP.
The distance $d$ between the window and the grating could be
smoothly varied from 0 to 45 mm. Also we had an option to change
the mutual polarization of the incident wave and the grating.

We used several 1D gratings (figure~\ref{Fig1a}a) having the same
outer size ($14.5\times 14.5$ cm) and the same spatial period
$D=2.26$ cm but with different slit widths $\Delta$: from 0.3 cm
to 1.96 cm. In addition to the ordinary grating with linear slits
we could also use a two-dimensional circular grating
(figure~\ref{Fig1a}b) having the same period and $\Delta=11.3$ cm.

\subsection{1D grating, short distance}

\begin{figure}[tbh]
\centering \scalebox{1.0}{\includegraphics{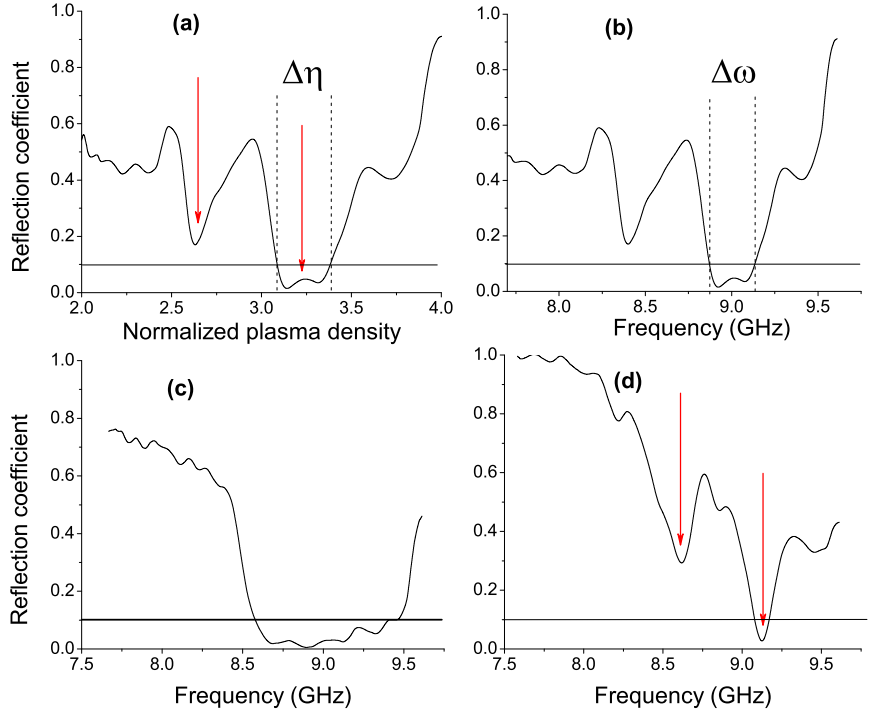}}
\caption{(Color online) Measured dependencies of the reflected
signal on the normalized plasma density $\eta=n_p/n_{c}$ for
gratings with different duty ratios $\alpha$. (a) $\alpha=0.13$,
$d=4$mm, dependence $R(\eta)$; (b) the same data presented as
dependence $R(\omega)$ using equation~(\ref{eq1}). The resonance
widths $\Delta\eta$ and $\Delta\omega$ are defined as the regions
where $R\leq 0.1$; (c)  $\alpha=0.5$, $d=3.5$mm,  (d)
$\alpha=0.75$, $d=4.5$mm. The level $R=0.1$ is indicated by the
horizontal line.} \label{Fig2}
\end{figure}

A set of movable 1D gratings with different periods $D$ and ``duty
ratios'' $\alpha=\Delta/D$ have been used in our experiments. When
the electric field of the incident wave was directed perpendicular
to the grating slits, the reflection reached zero for some
distance $d$ between the grating and the plasma surface,
independently of the value of $\alpha$. An example of the measured
dependencies of the reflected signal on the normalized plasma
density $\eta=n_p/n_c$ is shown in figure~\ref{Fig2}a. This
dependence of the reflection coefficient on the normalized plasma
density, $R(\eta)$, can be presented as the dependence of the
reflection coefficient on the incident wave frequency, $R(\omega)$
(figure~\ref{Fig2}b). Indeed, assuming that the angle of incidence
is small, $\theta\ll 1$, the resonance condition (\ref{eq1}) can
be presented as follows:
\begin{equation}\label{eq2}
\omega\sqrt{\left({\omega_0^2\over\omega^2}{n_p\over
n_{c0}}-1\right)/\left({\omega_0^2\over\omega^2}{n_p\over
n_{c0}}-2\right)}=ck_g.
\end{equation}
Here $\omega_0$ is the klystron operating frequency and $n_{c0}$
is the critical plasma density for this frequency. The solution of
equation~(\ref{eq2}) has the form:
\begin{equation}\label{eq3}
{\omega\over\omega_0}\equiv{f\over
f_0}=\sqrt{\left(\eta/2+\kappa_g^2\right)-\sqrt{\eta^4/4+\kappa_g^4}},
\end{equation}
where $\kappa_g=ck_g/\omega_0$ and $\eta=n_p/n_{c0}$.

Equation~(\ref{eq3}) allows presentation of the experimentally
measured dependence $R(\eta)$ for the fixed frequency $f_0=8.45$
GHz as the dependence of the reflection coefficient on the
incident wave frequency, $R(f)$, for a fixed plasma density
$n_p=n_{c0}(2\kappa_g^2-1)/(\kappa_g^2-1)$.

The dependencies $R(f)$ presented in Figs.~\ref{Fig2}b-d  for
gratings with different duty ratios $\alpha$ are characterized by
a set of peculiarities which can be explained theoretically in
what follows. First, the resonant plasma densities  corresponding
to the reflection minima were found close to the values
$n_{+1}\simeq 2.5 n_c$ and $n_{-1}\simeq 3.0 n_c$, which follows
from equation~(\ref{eq1a}), but did not coincide with them (see
figure~\ref{Fig2}a). The resonant frequencies (resonant plasma
densities) depend on the duty ratio $\alpha$ whereas this
dependence is absent in equation~(\ref{eq1a}). Second, the
dependencies $R(\omega)$ have a similar shape for the small
($\alpha=0.13$) and the large ($\alpha=0.75$) values of the
grating duty ratio $\alpha$. This shape is characterized by two
relatively narrow minima (marked by red arrows in
figure~\ref{Fig2}), one of which reaches zero. In contrast, in the
grating with the middle-value duty ratio ($\alpha=0.5$), the
dependence $R(\omega)$ is characterized by a very broad single
minimum. The range of frequency (plasma density) variation, which
corresponds to the small reflectivity, overlaps the regions of the
two  minima mentioned above.

\begin{figure}[tbh]
\centering \scalebox{1.1}{\includegraphics{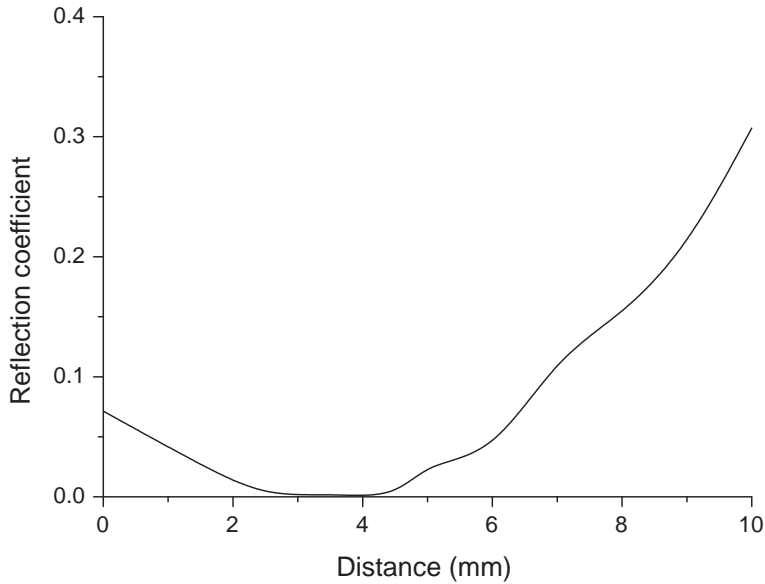}}
\caption{Measured dependence of the minimal reflected signal on
the distance $d$ between the grating and the plasma surface.}
\label{Fig3}
\end{figure}

The  dependencies $R(\omega)$ depicted in figure~\ref{Fig2}
correspond to the distances $d=d_c$ between the grating and the
plasma surface when the reflection drops to zero. These distances
$d_c$ are different for gratings with different duty ratio
$\alpha$: $d_c=d_c(\alpha)$. An example of the dependence of the
minimal value of the reflection coefficient on the distance $d$
for a given grating is shown in figure~\ref{Fig3}. The existence
of a certain distance $d$ when the reflected signal reaches zero
is the manifestation of the critical coupling effect.

\begin{figure}[tbh]
\centering \scalebox{1.0}{\includegraphics{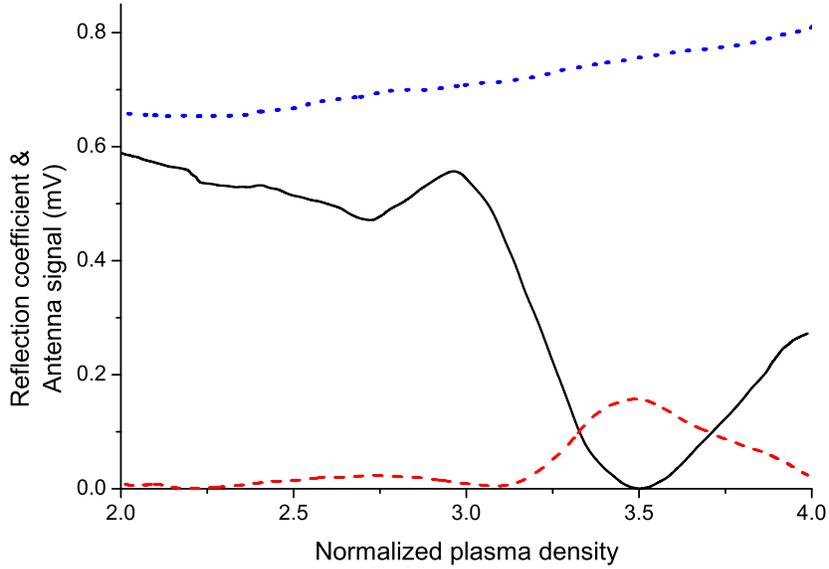}}
\caption{(Color online) Antenna signal (red dashed line) and
reflection coefficient (solid black and dotted blue lines). Solid
black line: the wave magnetic field is parallel to the grating
slits; dotted blue line: the wave electric field is parallel to
the grating slits. } \label{Fig4}
\end{figure}

The results described above have been obtained when the distance
$d$ is small, $d\sim 3\,-\,5$mm. In this case the SPP excitation
is responsible for the incident wave absorption. This is confirmed
by the detection by a miniature antenna of a significant rise of
the electric field under the plasma surface (see
figure~\ref{Fig4}). The absorption and the antenna signal
disappear when the wave electric field is directed along the
grating slits and the SPP can not be excited (dotted blue line in
figure~\ref{Fig4}).

\begin{figure}[tbh]
\centering \scalebox{1.0}{\includegraphics{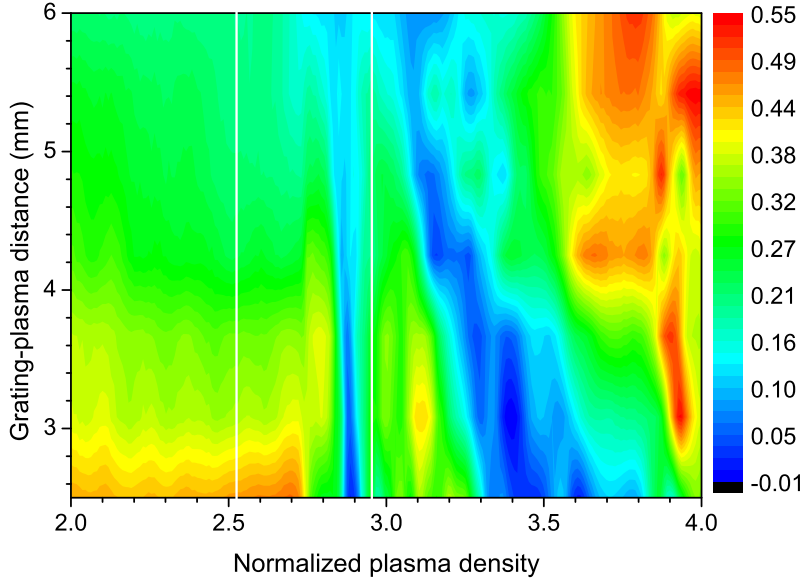}}
\caption{(Color online) Small grating-plasma distance $d$, grating
duty ratio $\alpha=0.13$. Experimental dependence of the
reflection coefficient $R$ (color map) on the normalized plasma
density $\eta$ and the distance $d$. Vertical white lines indicate
resonant plasma densities $\eta_{+1}$ and $\eta_{-1}$ calculated
using equation~(\ref{eq2}).} \label{Fig4a}
\end{figure}

The dependence of the reflection coefficient $R$ on the normalized
plasma density $\eta$ and the grating-plasma distance $d$ is shown
in figure~\ref{Fig4a}. The reflection is suppressed in two regions
that correspond to excitation of surface plasmon-polaritons
propagating in opposite directions. The deviation of the positions
of these experimentally determined regions from those calculated
using equation~(\ref{eq2}) (vertical white lines in
figure~\ref{Fig4a}) and the dependence of the positions on the
distance $d$, which is absent in equation~(\ref{eq2}), will be
discussed in what follows.

\subsection{1D grating, large distance}

\begin{figure}[tbh]
\centering \scalebox{1.0}{\includegraphics{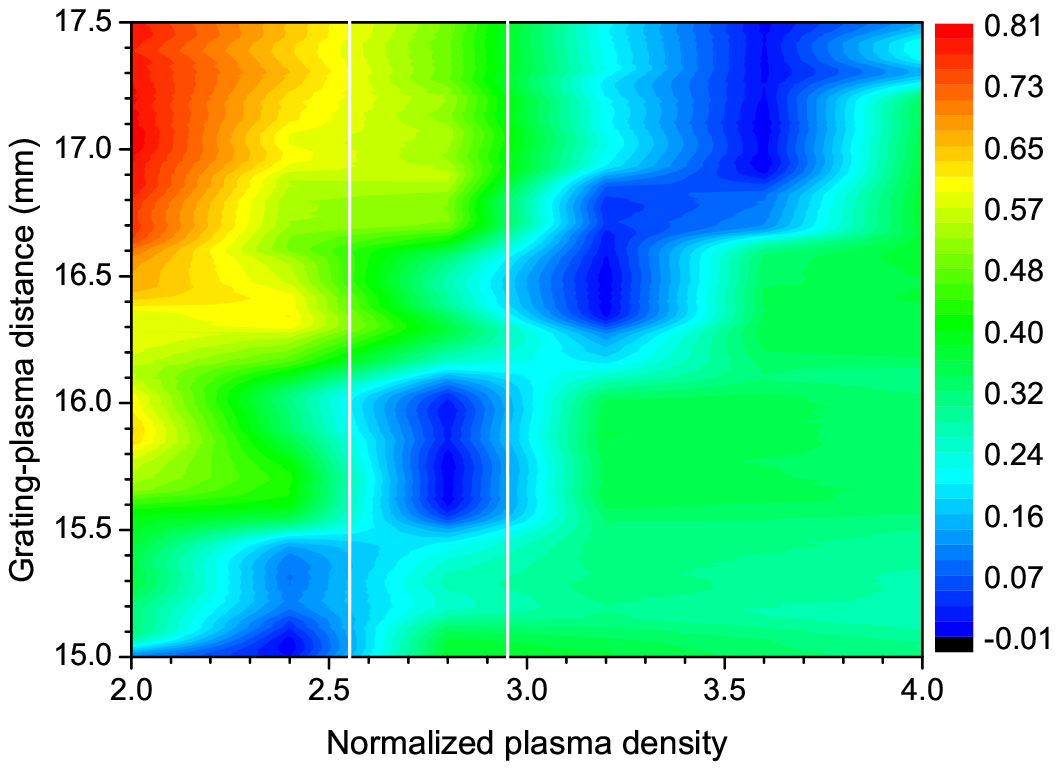}}
\caption{(Color online) Large grating-plasma distance $d$, grating
duty ratio $\alpha=0.13$. Experimental dependence of the
reflection coefficient $R$ (color map) on the normalized plasma
density $\eta$ and the distance $d$. Vertical white lines point to
theoretical values of the resonant plasma densities $\eta_{+1}$
and $\eta_{-1}$.} \label{Fig_5}
\end{figure}

\begin{figure}[tbh]
\centering \scalebox{1.1}{\includegraphics{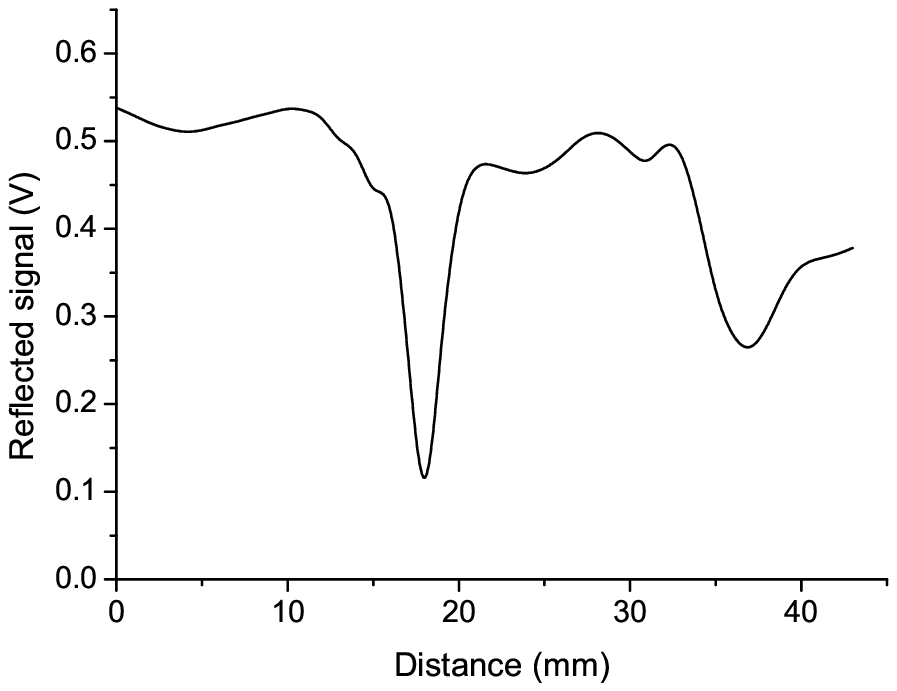}}
\caption{Reflection from the copper plate as a function of the
plate-grating distance.} \label{Fig_6}
\end{figure}

Another region of the reflection coefficient suppression has been
found at large distances $d$. A significant  decrease of the
reflection has been observed for gratings with low duty ratio
$\alpha$ only. Under the increase of the distance $d$, the
reflectivity minimum appears at low plasma density and moves fast
to  higher density (see figure~\ref{Fig_5}). The incident wave
absorption in this region of  distances $d$ is not connected with
the SPP excitation.  The absorption is caused by the excitation of
the resonator formed between the grating and plasma surface. It
was confirmed experimentally when a copper plate was used instead
of plasma surface. Results of this experiment are presented in
figure~\ref{Fig_6}.

Note that the distance $d$ that corresponds to the reflection
minimum coincides with $\lambda/2$ in the experiment with the
copper plate and is smaller than $\lambda/2$ in the experiments
with plasma. This distinction between the two experiments will
also be explained theoretically in the next section.

\subsection{2D circular grating}

\begin{figure}[tbh]
\centering \scalebox{0.5}{\includegraphics{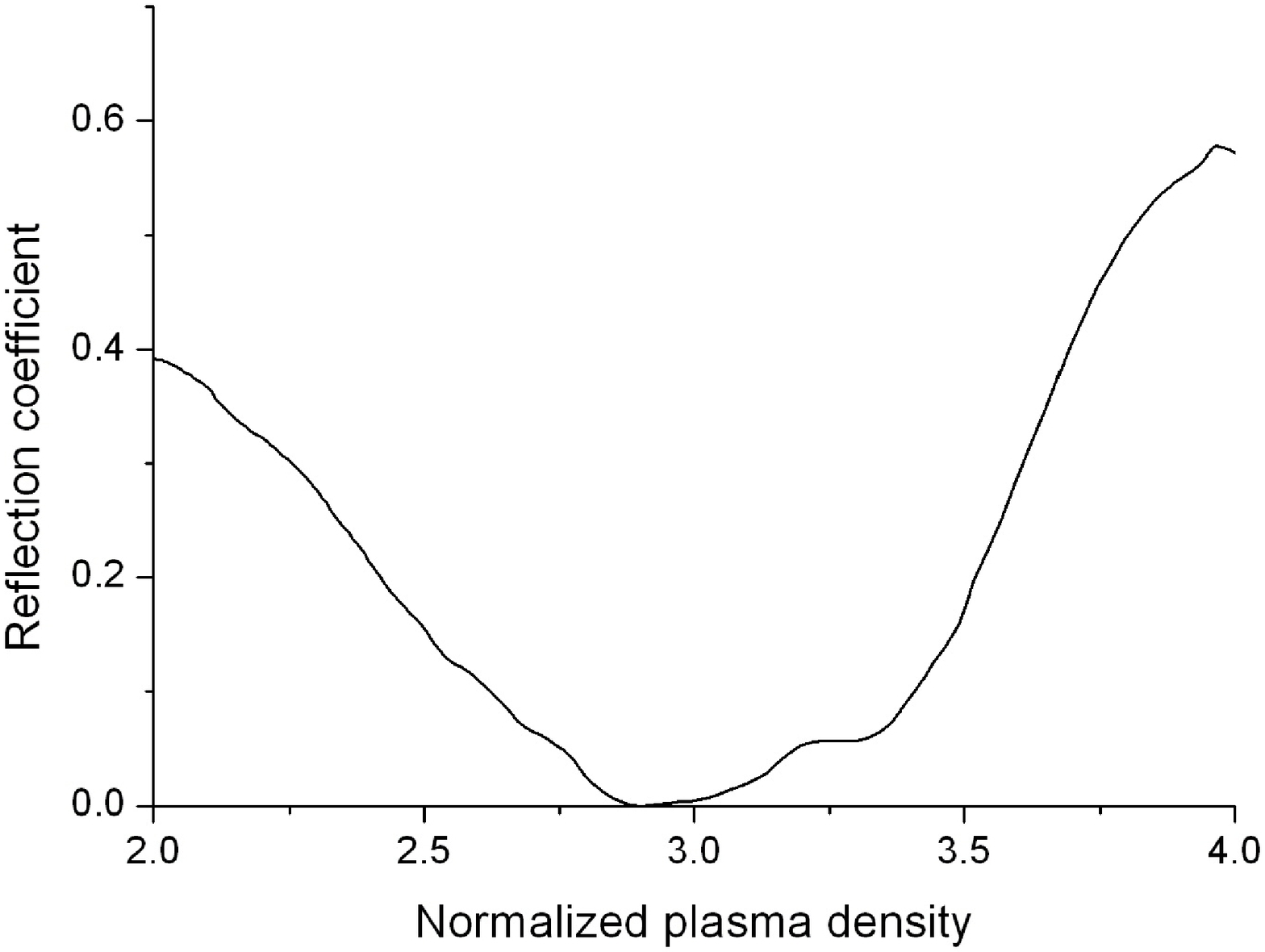}}
\caption{Circular grating. Dependence of the reflected signal on
the plasma density. } \label{Fig_7}
\end{figure}

The qualitative analysis presented in section II has been
successfully confirmed experimentally. In figure~\ref{Fig_7} the
dependence of the reflected signal on the plasma density is shown.
The reflection is characterized by a broad minimum and reaches
zero at the resonant plasma density.  Recalculating the dependence
$R(\eta)$ as the dependence $R(f)$, one can see that the
reflection coefficient does not exceed 10\% in the frequency band
$f\in (8.4 - 9.15)$GHz, i.e. the relative half-width $\Delta f/f$
reaches 4\%. This value essentially exceeds the relative
half-width $\Delta f/f\simeq 0.3$\% obtained in the Ref.~22 with
the use of a doubly periodic grating with square symmetry. This
difference can be explained qualitative in the following way.

The circular grating can be conventionally characterized by a
radial reciprocal vector $\mathbf{k}_g(\phi)$ whose direction is
varied continuously with the angular coordinate $\phi$. When the
angle of incidence differs from zero ($8^\circ$ in the
experiment), the scattered field is characterized by a continuous
set of wave vectors
$\mathbf{k}(\phi)=\mathbf{k}_0+\mathbf{k}_g(\phi)$. It is worth to
note once more that in this case the reflection coefficient is
independent of the wave polarization.

\section{Theoretical model}

\begin{figure}[tbh]
\centering \scalebox{1.0}{\includegraphics{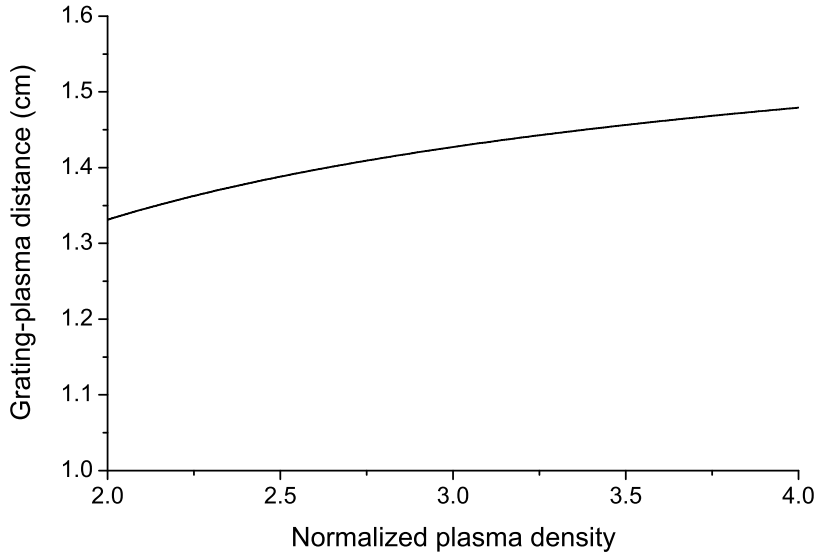}}
\caption{Dependence of the resonant distance $d_{res}(\eta)$ on
the normalized plasma density $\eta$ [solution of the dispersion
equation (\ref{eq4})]. } \label{Res2_distance}
\end{figure}

The diffraction grating is semitransparent for propagating waves,
therefore the space between the overdense plasma (which reflects
propagating waves) and the grating forms a resonator which will be
marked below as a resonator-2 whereas SPP will be marked as
resonator-1. An eigenfrequency of the resonator-2 is determined by
the condition that the wave accumulates the phase by multiples of
$2\pi$ along the round trip between the grating and the plasma
surface. It is necessary to note that reflection from a plasma
whose density $n_p$ is large but finite, is accompanied by a
non-zero phase shift $\delta\phi(n_p)$. As a result, the distance
$d_{res}$ between the grating and the plasma for which the wave
frequency coincides with the resonator-2 eigenfrequency depends on
the plasma density:
\begin{equation}\label{eq4}
d_{res}=(k_0\cos\theta)^{-1}\arctan[-(\eta-1)^{-1/2}].
\end{equation}
The resonant distance $d_{res}$ tends to $\lambda/2$ (where
$\lambda$ is the wavelength) asymptotically when the plasma
density tends to infinity, $d_{res}\rightarrow\lambda/2$ when
$n_p\rightarrow\infty$. This explains why the experimentally
measured distance which corresponds to the reflection minima is
smaller than $\lambda/2$ and depends on the plasma density as it
is shown in figure~\ref{Fig_5}. The dependence in
equation~(\ref{eq4}) of the resonant distance on the plasma
density is depicted in figure~\ref{Res2_distance}. The density in
this figure is varied in the range $\eta=n_p/n_c\in(2,4)$ that
corresponds to the experimental conditions.

Resonator-2 also possesses radiative and dissipative Q-factors.
The first one is determined by the grating transparency, i.e. by
the grating duty ratio $\alpha$, whereas the second one is
determined by the dissipation in the plasma skin layer.

When the distance $d$ between the grating and the plasma surface
coincides with $d_{res}$ (or the wave frequency coincides with the
resonator-2 eigenfrequency) and the critical coupling condition is
satisfied, the reflection coefficient vanishes. This mechanism of
the wave absorption by the overdense plasma is not connected with
the excitation of surface waves.

Thus, there are two separate domains of the system parameters
where reflection can be strongly suppressed. They are connected
with excitation of either resonator-1 or resonator-2. The critical
coupling condition, ${\rm Q}_{diss}={\rm Q}_{rad}$, can be
achieved for  resonator-1 by a proper fitting of the distance $d$
which determines the radiative Q-factor. In contrast, this can not
be done for resonator-2, because the distance $d$ determines not
the radiative Q-factor but the eigenfrequency of this resonator.
The radiative Q-factor of resonator-2 depends on the grating
transparency, i.e. on the grating duty factor $\alpha$  and the
mutual orientation of the wave polarization vector and the grating
reciprocal vector $\mathbf{k}_g$.

The above-mentioned peculiarities of the experimental results can
be explained using the following simple model of coupled
resonators, two resonators-1 (when the angle of incidence is not
equal to zero) and one resonator-2. The monochromatic incident
wave penetrating through a semitransparent wall (diffraction
grating) excites resonator-2. The scattering of the resonator-2
wave field  on the grating produces the evanescent waves which
excite the SPPs (two resonators-1) at the plasma surface. This
process can be modeled by the following equations:

\begin{eqnarray}\label{A1}
\ddot{\psi_+}+\omega^2_+(\eta)\psi_++\gamma_1\dot{\psi_+}=q(d)\psi_2,\nonumber\\
\ddot{\psi_-}+\omega^2_-(\eta)\psi_-+\gamma_1\dot{\psi_-}=q(d)\psi_2,\nonumber\\
\ddot{\psi_2}+\omega^2_2(\eta,d)\psi_2+\gamma_2(d)\dot{\psi_2}=q(d)\psi_++q(d)\psi_-+fe^{-i\omega
t}.
\end{eqnarray}

Here $\omega_{\pm}$ are the resonators-1 eigenfrequencies,
$\omega_2$ is the resonator-2 eigenfrequency, $\gamma_{1,2}$ are
the corresponding resonators dissipation coefficients, $f$ is
proportional to the incident wave amplitude, and $q$ is the
coupling coefficient which describes mutual transformation of the
propagating and evanescent waves on the grating surface. The
functions $\psi_\pm$ and $\psi_2$ represent the resonators-1 and
resonator-2 fields, respectively.

The eigenfrequaencies $\omega_\pm$ and $\omega_2$ are defined by
the dispersion equations (\ref{eq1}) and (\ref{eq4}) which should
be solved for $\omega$ as a function of the normalized plasma
density $\eta$ and the grating-plasma distance $d$. Note, that a
real solution $\omega_2(\eta, d)$ of equation~(\ref{eq4}) exists
only when $k_0 d>\pi/(2\sqrt{\eta})$. The coupling coefficient $q$
depends on the distance $d$  as $q=q_0\exp(-dk_p)$, where
$k_p=\sqrt{k_g^2-k_0^2}$ is the evanescent wave spatial decrement.
The dissipation coefficient $\gamma_1$ is associated with the
plasma electron collision frequency $\nu$:
$\gamma_1\approx\nu/\omega$. The coefficient $\gamma_2$ is defined
by the grating transparency, i.e., by the grating duty ratio
$\alpha$, and the field leaking through the open lateral surface
of resonator-2.

\begin{figure}[tbh]
\centering \scalebox{1.0}{\includegraphics{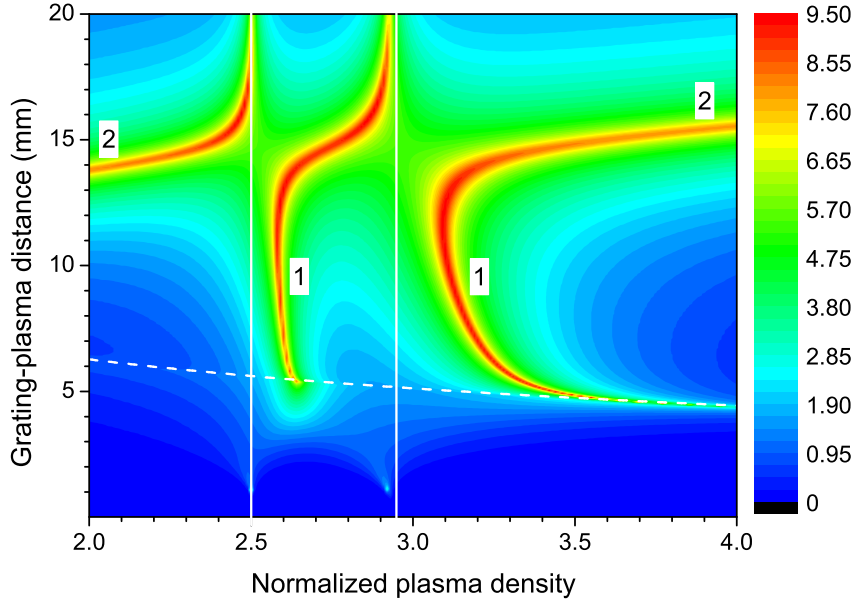}}
\caption{(Color online) Theoretical dependence of the
electromagnetic energy density $W$ on the normalized plasma
density $\eta$ and the distance $d$. The resonant plasma density
[solution of the dispersion equation (\ref{eq1})] is shown by
white vertical line. The real solution $\omega_2(\eta, d)$ exists
above the dashed white line. 1 -- regions where dissipation is
connected with the resonator-1 excitation (SPP excitation), 2 --
regions where the dissipation is related to the resonator-2
excitation. } \label{Theor2}
\end{figure}

The suppression of the incident wave reflection is associated with
an enhanced dissipation rate $P$ of the electromagnetic energy in
the plasma. This dissipation, in turn, is proportional to the
energy density of the electromagnetic field in the plasma skin
layer, i.e., it is proportional to the energy density $W$ of the
fields of all three resonators: $P\propto
W=|\psi_+|^2+|\psi_-^2|+|\psi_2|^2$. The energy density $W$ is
maximal near the resonant frequencies of the system which are
shifted from their non-perturbed values $\omega_+$, $\omega_-$,
and $\omega_2$ due to the coupling between the resonators. Thus,
the solution of Eqs.~(\ref{A1}) allows determination of the
regions in the $\eta, d$-plane where the reflection coefficient
$R$ can be small. The  relative distribution of the
electromagnetic energy $W(\eta, d)$ obtained by numerical solution
of equation~(\ref{A1}) is shown in figure~\ref{Theor2} as a color
map. The reflection is suppressed in the red-yellow-colored
regions. Solutions of equation~(\ref{A1}) with various values of
the parameters $\gamma_1$, $\gamma_2$, and $q_0$ showed that the
position of the regions where the energy density $W$ is large
depends mainly on the value of the coupling coefficient $q_0$,
whereas the dissipative coefficients $\gamma_{1,2}$ determine the
regions width. Some parameters like the operating frequency
$\omega$ and the grating period $D$ are exactly determined by the
experimental conditions, while other parameters can be estimated
using the plasma parameters and the geometry of  resonator-2 as
follows: $\gamma_1\sim\nu/\omega_p\sim 10^{-3}$,
$\gamma_2/\omega\sim 10^{-1}$.

Let us compare now the experimental results presented in
figures~\ref{Fig4a} and \ref{Fig_5} with the theoretical results
depicted in figure~\ref{Theor2}.
The minima of the reflection coefficient $R$ coincide with the
maxima of $P$.  Figure~\ref{Theor2} shows, first, that for small
distances $d$ when the reflection suppression is associated with
the SPP excitation, the plasma density corresponding to the
reflection coefficient minima is higher than the resonant plasma
densities defined by equation~(\ref{eq1}). This is in agreement
with the experimentally measured dependencies of the reflection
coefficient on the plasma density. Second, upon increasing the
distance $d$, the difference between the resonant densities
$\eta_{\pm 1}$ defined by equation~(\ref{eq1}) and the densities
corresponding to the reflection minima decreases both in the
experiment and in the theory. Third, the dependence of the lower
resonant density on the distance is much weaker than it is for the
higher resonant density, both in figures~\ref{Fig4a} and
\ref{Theor2}.

As regards to the large distances, the experimental as well as the
theoretical data demonstrate the same behavior: the resonant
distance $d_{res}$ where the reflection is small, depends weakly
on the normalized plasma density and grows when the density
increases (compare figure~\ref{Fig_5} and regions ''2'' in
figure~\ref{Theor2}). Moreover, two gaps in the reflection minima
which are located in the vicinity of the unperturbed resonant
densities $\eta_{\pm 1}$ (white lines in the figures), are clearly
visible in the experimental and the theoretical figures.

Thus, the simple theoretical model of coupled resonators explains
qualitatively all the peculiarities of the experimentally measured
dependencies.

\section{Summary}

We have shown both experimentally and theoretically that the range
of parameters where surface plasmon-polaritons are effectively
excited by a freely propagating electromagnetic wave can be
essentially broadened due to combination of two factors: a
non-zero angle of incidence and a two-dimensional circular
diffraction grating. The usage of axially-symmetrical grating
makes the SPP excitation entirely independent of the wave
polarization. In our experiments we used a wave beam whose
transversal cross section coincided with the grating aperture. The
2D superlattice composed of circular gratings used in the
experiments is able to utilize completely a wide wave beam for SPP
excitation. This statement can be verified in the infrared or
visible frequency range with a solid state plasma.

There are two sets of problems where the total absorption of
electromagnetic waves is a desired result. The first one is
cloaking -- making an object invisible in a reflected signal
(making invisible for radar, for instance). There is no difference
for this purpose where the incident wave energy disappears. For
the second set of problems (plasma heating, enhancement of
photovoltaic cells efficiency, etc.) the dissipation mechanism
plays the key role. Depending on the aim, either localized or
propagating surface waves can be used the most effectively.
Indeed, localized plasmon-polaritons are eigenmodes of
subwavelength resonators (voids \cite{Teperik}, silver
microparticles \cite{Pillai}) and dissipation of an incident wave
occurs mainly due to the energy loss in these resonators. Part of
the wave energy can be used for intensification (due to
enhancement of electromagnetic fields in the vicinity of these
resonators) of some processes (photoeffect, for instance) in a
substrate. In contrast to localized modes, propagating
plasmon-polaritons provide spatially homogeneous coupling with the
substrate (or even being eigenmodes of the substrate interface, as
it occurs with plasma) and the energy dissipation can be caused
mainly by processes in the substrate.

Thus, the merits of localized plasmon-polaritons are low
sensitivity to the electromagnetic  wave angle of incidence and
polarization. A disadvantage is their relatively weak and
spatially inhomogeneous coupling with a substrate. When the goal
is reflectivity suppression only, this disadvantage is of no
significance. On the contrary, propagating (delocalized)
plasmon-polaritons are sensitive to the wave polarization and
angle of incidence, but can be rather strongly coupled with the
substrate. The results presented in this paper show how this
sensitivity can be essentially suppressed.

\end{document}